# Phase Boundaries, Isotope Effect and Superconductivity of Lithium Under Hydrostatic Conditions

Stefano Racioppi,[1] Iren Saffarian-Deemyad,[2,*] William Holle[2,*], Francesco Belli,[1] Richard Ferry,[3] Curtis Kenney-Benson,[3] Jesse S. Smith,[3] Eva Zurek,[1] Shanti Deemyad[2]

[1]Department of Chemistry, State University of New York at Buffalo, Buffalo, New York 14260-3000, USA.
[2]Department of Physics and Astronomy, University of Utah, Salt Lake City, Utah 84112, USA.
[3]High Pressure Collaborative Access Team, X-ray Science Division, Argonne National Laboratory, Argonne, IL, 60439 USA

*These authors contributed equally to the work

**Abstract**
We present theoretical and experimental studies of superconductivity and low temperature structural phase boundaries in lithium. We mapped the structural phase diagram of $^6$Li and $^7$Li under hydrostatic conditions between 5 – 55 GPa and within the temperature range of 15 – 75 K, observing the FCC → hR1 → cI16 phase transitions. $^6$Li and $^7$Li show some differences at the structural boundaries, with a potential shift of the phase boundaries of $^6$Li to lower pressures. Density functional theory calculations and topological analysis of the electron density elucidates the superconducting properties and interatomic interactions within these phases of lithium.

**Introduction**
A microscopic understanding of the evolution of the electronic structures of elemental solids upon compression remains a grand challenge in high pressure science. Lithium is the lightest element to become a superconductor at ambient pressure, with a critical transition temperature, $T_c$, of 4 mK [1–3]. Quantum effects in lithium result in a pressure-temperature ($P – T$) phase diagram that is intricate [4–6], including transitions to multiple superconducting phases, achieving a $T_c$ as high as ~15 – 20 K, the highest reported among elements below 100 GPa [7–9]. Beyond this range, lithium transforms into a semiconductor around 80 GPa [10] and into a metal at pressures exceeding 120 GPa [11]. Despite decades of intensive investigations aiming to establish a general consensus on the sequence of pressure induced phase transitions in lithium, there is notable variation in the boundaries of these phases across different studies, possibly due to differences in hydrostatic conditions and measurement techniques.

One of the primary disagreements in the lithium phase diagram is in the pressure range of 30 – 50 GPa, particularly at low temperatures, encompassing the majority of lithium's superconducting region. In this range, various studies report two phase transitions in lithium: FCC (*Fm-3m*) → hR1 (*R-3m*) → cI16 (*I-43d*). However, significant discrepancies exist in the boundaries of these phases [5].

The hR1 phase, in particular, appears within a narrow pressure window, seldom as a pure phase, and with ambiguous temperature dependencies [4–6,12,13]. Initially observed by Hanfland *et al.* between 39 – 42 GPa at 180 K, subsequent measurements proposed alternative $P – T$ boundaries. For example, Guillaume *et al.* [4] noted the disappearance of the rhombohedral phase below 120 K, suggesting a direct transformation from FCC to cI16. Conversely, Matsuoka *et al.* [13] observed hR1 down to 25 K, potentially persisting up to 46 GPa, although the presented data lacks critical structural information to define boundaries or fully support the reported structures. A more recent study by the same group [11], conducted at 50 K, reports a sequence of structural phase transitions with the observation of hR1 between 39 and 44 GPa. Frost *et al.*, however, suggested the observation of hR1 at 41 GPa and 255 K, despite their reported powder X-ray diffraction pattern aligning better with the cI16 phase [6].

The pressure range where the FCC → hR1 → cI16 phase transformations occur exhibits a peculiar behavior in the superconducting properties [7–9,11]. Measurements show that between 20 – 30 GPa, when lithium is stable in the FCC phase, superconductivity is quickly enhanced by compression, reaching $T_c$ ~ 15 K under hydrostatic conditions [7–9]. However, at higher pressures, the superconducting trend changes course twice, initially decreasing to ~ 40 – 45 GPa and then increasing again until it suddenly vanishes above ~ 70 GPa [9–11]. Theoretical calculations emphasize the importance of soft phonon modes near the transition pressure for the enhancement of the superconductivity in the FCC phase [14–16]. However, due to the lack of structural measurements in the superconducting region, none of the theoretical analyses of lithium's superconductivity beyond 30 GPa have been based on experimental data.

One of the primary open questions about the phase diagram of lithium, directly impacting our understanding of its superconducting behavior, is the structural boundaries of its hR1 phase at temperatures comparable to the $T_c$. In this work, we determine these specific phase boundaries under hydrostatic conditions. We complement our study with density functional theory (DFT) calculations coupled with topological analysis of the electron density to elucidate the thermodynamics and electronic structure changes along the FCC → hR1 → cI16 phase transformations. Our theoretical analysis helps explain the superconducting phase diagram of lithium and provides a coherent picture of the superconducting properties of this element.

**Result and Discussion**
The phase diagram of lithium is determined under nearly hydrostatic conditions with helium as a pressure transmitting medium (PTM) in a diamond anvil cell (Figure S1a). Our measurements covered the $P – T$ range between 5 – 55 GPa and 15 – 75 K, focusing on the phase transition sequence FCC → hR1 → cI16 (Figure 1a, S1b). Synchrotron x-ray diffraction (XRD) data were collected at the 16-ID-B beamline, HPCAT of the Advanced Photon Source (APS), Argonne National Laboratory (for further experimental detail, see Section S1 in the Supplemental Material). We detected pure hR1 in a very narrow pressure range (only at a couple of points: ~36.4 GPa at 75 K and ~40.6 GPa at 15 K, Figure 1, S1b). Otherwise, hR1 is predominantly detected as a mixture with FCC (34.5 – 36 GPa at 75 K and 36 – 40 GPa at 15 K) or with cI16 (40 – 41.5 GPa at 75 K). This confirms that hR1 persists to the lowest measured temperatures as previously observed at higher temperatures [5]. Furthermore, powder diffraction data indicates a shift in the pressure boundaries of hR1 to lower values at 75 K, persisting over a broader compression range. In Figure 1b, we compare our FCC → hR1 → cI16 structural phase boundaries, with the superconducting phase diagram of lithium measured previously under similar experimental conditions, and our theoretical estimates of $T_c$ (detailed below) of the corresponding structures. The comparison shows good qualitative agreement with the trend experimentally observed for the oscillating $T_c$ behavior of lithium at the FCC → hR1 → cI16 phase boundaries (See Section S2 in the Supplemental Material for further details).

In addition, we detected a new peak at ~10.93 ° in the range of stability of hR1, which had the highest intensity when the phase was pure (Figure S1c). The relatively low intensity of this peak makes it plausible that it has been overlooked in previous studies. We suggest that this peak might be indexed as the (100), which corresponds to a systematic absence in the space group *R-3m*. Nonetheless, this reflection can be, in principle, activated upon slight distortions of the unit cell (See Table S1), suggesting either hysteresis or deviation from the ideal geometry under compression.

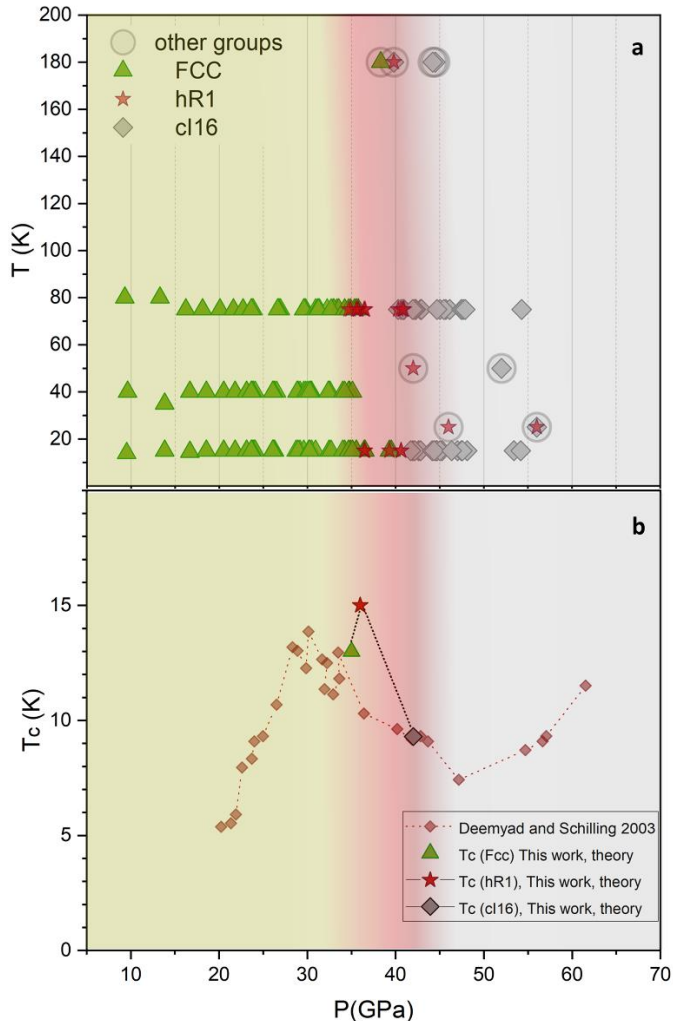

**Figure 1.** (**a**) Phase diagram of lithium under hydrostatic conditions probed in this work and compared with previous studies (Refs. [5,11,13]). (**b**) The superconducting phase diagram of $^7$Li, obtained from measurements conducted under hydrostatic conditions in helium as the PTM (Ref. [8]). The diagram also includes the calculated values of the superconducting critical temperature at the corresponding pressures determined in this study.

The difference of one neutron in the nucleus leads to ~14 % change in the atomic mass of stable lithium isotopes ($^6$Li and $^7$Li). This difference significantly contributes to both relative quantum fluctuations and phonon dynamics, which was suggested to influence the superconducting behavior of lithium isotopes under pressure. [17] We conducted comparative measurements of the phase diagram of the two lithium isotopes up to ~40 GPa and at a temperature below 15 K along an isothermal compression path and under very similar experimental conditions (Figure 2). $^6$Li and $^7$Li exhibit different behavior already at mild compression ($P$ < 5 GPa) [17], displaying variations in their martensitic transition temperature. In contrast to $^7$Li, the BCC structure of the $^6$Li samples remains stable relative to the martensite down to the lowest measured temperatures for pressures up to around 2 GPa during isochoric cooling.

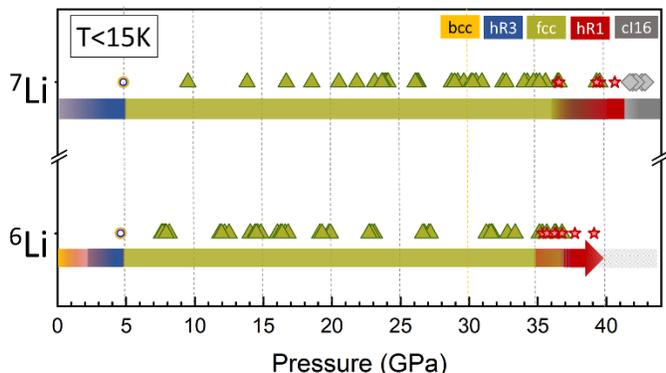

**Figure 2.** The structural phase boundaries of stable isotopes of lithium at 15 K. Symbols and colors show various phases: Yellow and blue open circles (BCC and hR3 respectively), olive triangles (FCC), red stars (hR1) and gray diamond (cI16). All measurements were carried out using helium as a pressure-transmitting medium. In the case of $^6$Li, the onset of the hR1 phase was observed at densities that correspond to approximately 1 GPa lower pressure in $^7$Li, and a pure phase was also noted at densities corresponding to slightly lower pressures compared to $^7$Li. The gray shaded area for $^6$Li represents the area where we do not have experimental data.

Our measurements here reveal that the $^6$Li transformation from FCC → hR1 occurs at densities that correspond to ~1 GPa lower pressures compared to $^7$Li. In the comparison of the phase boundaries of the two isotopes, we relied on a direct assessment of the unit cell size of the sample at the phase boundaries. For example, the Bragg peaks of $^6$Li at the end of the FCC phase are at lower angles, or a larger unit cell volume compared to $^7$Li. In our measurements on $^7$Li, the calibration was made based on pressure from ruby florescence and the volume of the sample portion at the same location. The calculated pressures in Figure 2 are estimated based on this calibration, assuming that the bulk moduli for both lithium isotopes at its FCC structure are the same. It is notable that even at ambient conditions the $^6$Li unit cell is slightly larger in volume due to larger nuclear zero-point effects [18]. Though using the same compressibility for two isotopes might introduce small errors, it would not change the conclusion regarding the isotope dependent differences observed here. Since $^6$Li is the lighter of the two isotopes, effects driven by quantum vibrations are expected to be more relevant, affecting the phase boundaries. This aspect will be discussed in more detail in the next paragraph.

Having established that hR1 exists within a very narrow pressure range, influenced by the isotopic weight of the metal, the next question to address is "why?". Past theoretical analyses proposed hR1 as the ground state phase of lithium at $P \sim 40$ GPa after interpolation of the DFT enthalpies ($H = E + PV$) at various pressures [19,20]. More recent computations have shown that the enthalpy of hR1 is never simultaneously lower than that of FCC and cI16 [16]; a result reproduced also by our DFT calculations (Figure S3a). However, we additionally noticed some differences introduced by changing the level of theory from the generalized gradient approximation (GGA) to meta-GGA [21–23] (see Section S3 in the Supplemental Material for further details). If we compare the enthalpy (static lattice at 0 K) of hR1 relative to FCC, PBE predicts the two phases to be isoenthalpic up to 42 GPa; on the other hand, R$^2$SCAN-L, a meta-GGA exchange correlation functional, finds that hR1 becomes thermodynamically preferred at 32 – 33 GPa, though by no more than a few meV/atom. The similarity in the enthalpy between these two phases is somehow expected, since hR1 is simply an FCC lattice that has been distorted along the diagonal of the cubic unit cell. The cI16 structure of lithium, however, becomes the enthalpically most stable phase already between 30 and 34 GPa, depending on the DFT functional. Therefore, the range of pressures considered here finds three phases very close in energy, and two of them (FCC and hR1) are also structurally similar.

Accounting for quantum lattice effects, such as zero-point energy corrections (ZPE), and temperature factors (within the PBE approximation), helps to increase the resolution at the FCC – hR1 boundaries. At 31 GPa, both phases are calculated to be dynamically stable in the harmonic approximation. However, at 33 GPa, the FCC phase gains an imaginary mode along the Γ → R path (~ -20 cm$^{-1}$, see Figure S4). The ZPE has a destabilizing effect on both phases, but to a greater extent on FCC, favoring the hR1 phase even at 0 K (Figure S3b). Additionally, when finite temperature effects are included, the Gibbs free energy favors hR1 even more (ΔG was estimated also at 33 GPa by assuming the small imaginary frequency of FCC as real).

An even clearer understanding of the FCC – hR1 phase boundary can be obtained considering quantum nuclear effects (QNEs). QNEs introduce anharmonic corrections in the phonon spectra, which may lead to phonon renormalization, especially for light elements such as lithium [24]. QNEs were calculated through the Stochastic Self-Consistent Harmonic Approximation [25] (SSCHA, see Section S4 in the Supplemental Material). Accounting for the anharmonicity of the atomic motions influences the lower optical phonon branches of both phases (Figure S5). FCC becomes dynamically unstable at around 38 GPa, while hR1 is observed to be dynamically stable between 38 and 39 GPa at 0 K, in good agreement with the experimental observations. Moreover, the SSCHA also affects the hR1 mode near the M-point, which was computed as being soft within the harmonic approximation, but undergoes a severe renormalization after accounting for the anharmonicity.

The theoretical FCC – hR1 boundaries are evidently susceptible to the level of theory, and the inclusion of QNEs. Nonetheless, in all cases, hR1 appears as a structural distortion of FCC, which survives only in a narrow range of pressure conditions, and is often mixed with the cubic phase (Figure 1-2, S1b). Therefore, hR1 might be interpreted as a local minimum in the path of the potential energy surface connecting FCC to cI16, probably characterized by low FCC ↔ hR1 energy barriers due to the structural similarity between the two phases.

To understand the origin of the "roller-coaster" profile of *T*c versus pressure measured around 30 – 40 GPa, we calculated the superconducting properties for lithium FCC, hR1 and cI16 at the pressures of 35 GPa, 36 GPa and 42 GPa, respectively, within the harmonic approximation (see Section S3 in the Supplemental Material for further details). The calculated values of *T*c (using the Allen Dynes modified McMillan equation), $\omega_{ln}$ (the logarithmic average frequency), and $\lambda$ (electron phonon coupling) are provided in Figure 3 and Table S2, together with the phonon band structures, the Eliashberg spectral function and $\lambda(\omega)$.

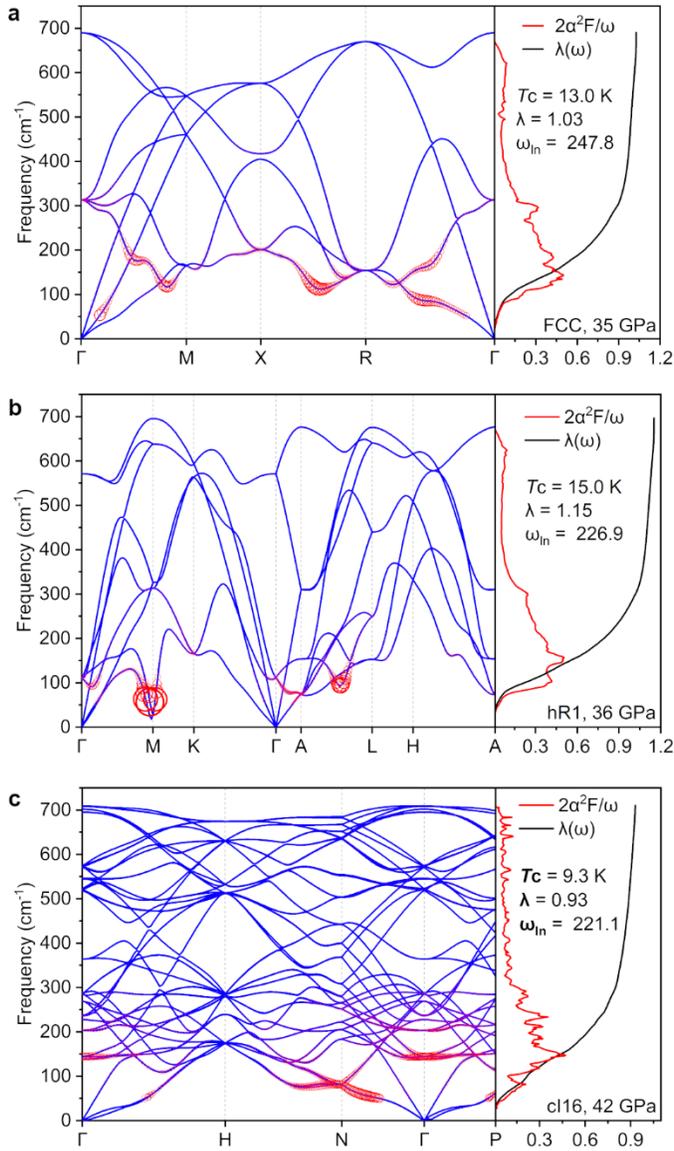

**Figure 3.** Phonon band structures, Eliashberg spectral function, $2\alpha^2 F(\omega)/\omega$, integrated electron-phonon coupling, $\lambda$, critical Temperature ($T$c), and logarithmic average phonon frequency $\omega_{ln}$ calculated for (**a**) FCC at 35 GPa, (**b**) hR1 at 36 GPa and (**c**) cI16 at 42 GPa. The diameter of the red circles in the phonon band structures scales with the electron-phonon coupling contribution. The Coulomb repulsion term $\mu^*$ was set equal to 0.17, in accordance with previous works [16].

The estimated $T$c equals 13.0 K at 35 GPa for FCC, 15.0 K at 36 GPa for hR1 and 9.3 K at 42 GPa for cI16, well in agreement with past experimental measurements [8,9] (see Figure 1b and S2). Bazhirov *et al.* [15] were the first to report that the change in $T$c, observed in Li-FCC, was triggered by the softening of certain phonon modes as a result of an incipient structural instability. Figure 3a shows analogous features, with the largest electron phonon coupling given by soft phonon modes along the Γ → M, X → R and R → Γ paths. On the other hand, Li-hR1 at 36 GPa, which corresponds to a distortion of the FCC lattice, shows two soft modes with particularly strong electron-phonon coupling strength: one near the M point and one along the A → L path. The SSCHA analysis showed that the soft phonon mode at M is renormalized after accounting for the anharmonic motions (Figure S5); however, the mode along A → L, which coincides with the atomic motions leading to the FCC ↔ hR1 distortion, is retained. The electron-phonon analysis suggests that the peak in the $T$c – $P$ graph measured by Deemyad and Schilling [8] (See also Figure 1b and S2) might be assigned to the emergence of the hR1 phase. The Eliashberg spectral function of cI16, instead, differs substantially from the previous two phases (Figure 3c). The electron-phonon coupling constant decreases by almost ~20% compared to hR1, causing the drop of $T_c$. Therefore, the measured negative slope of the superconducting critical temperature that leads to a minimum at ~42 GPa, can be assigned to the transition from hR1 to cI16.

The above analysis explains and reproduces the observed trends in the measured $T_c$ for FCC and hR1, as well as its drop within cI16. However, can the electronic structure of these phases give us more insight on the structural – superconductivity relationship within lithium? In this context, the calculated density of states (DOS) at the Fermi level ($g(E_F)$), plots of the Electron Localization Function (ELF) [26,27] and Quantum Theory of Atoms in Molecules (QTAIM) [28] descriptors, become useful tools [29,30] (See Section S5 in the Supplemental Material for further details).

In Figure 4a-c, we report the DOS of the three phases, together with their orbital projections. The almost indistinguishable projected DOS and $g(E_F)$ values of FCC and hR1 testifies to their structural similarity, showing that both valence states are dominated by the 2p orbitals. On the other hand, the occupied valence DOS of cI16 has a smaller dispersion, indicative of a larger degree of localization, which is also characterized by the formation of a pseudo-gap. The pseudo-gap was calculated also by Hanfland *et al.*, and originally attributed to a Peierls distortion of the BCC lattice [5]. These features of cI16 suggest the electronic structure differs from the two other phases. Therefore, let us shift our attention to an analysis of its descriptors: loci with ELF values > 0.8 are present in all three systems (Figure S6a-c and Figure 4a-c inset), enclosing regions of space characterized by large Pauli repulsion. Alkali metals possess only one valence electron, used to form metallic bonds. Each of these valence electrons is surrounded by a depletion in the probability of finding another electron with the same spin, a Fermi hole, which arises from the Pauli exclusion principle. Consequently, these depletions give rise to large values in the ELF in regions of space corresponding to the singly occupied valence orbitals of the constituent atoms. [29–34]. Similarly, a QTAIM analysis in solid phases finds non-nuclear maxima (NNM), located at the center of the ELF's isosurfaces [32]. The charge, Q, integrated within the region of space surrounding the NNM (Bader's basin), the electron density at the NNM, $\rho_{NNM}$, and the negative of the Laplacian of the electron density at the same point, $-\nabla^2\rho_{NNM}$, quantify the amount of electron density at these critical points, as well as their degree of localization (Figure 4d-f).

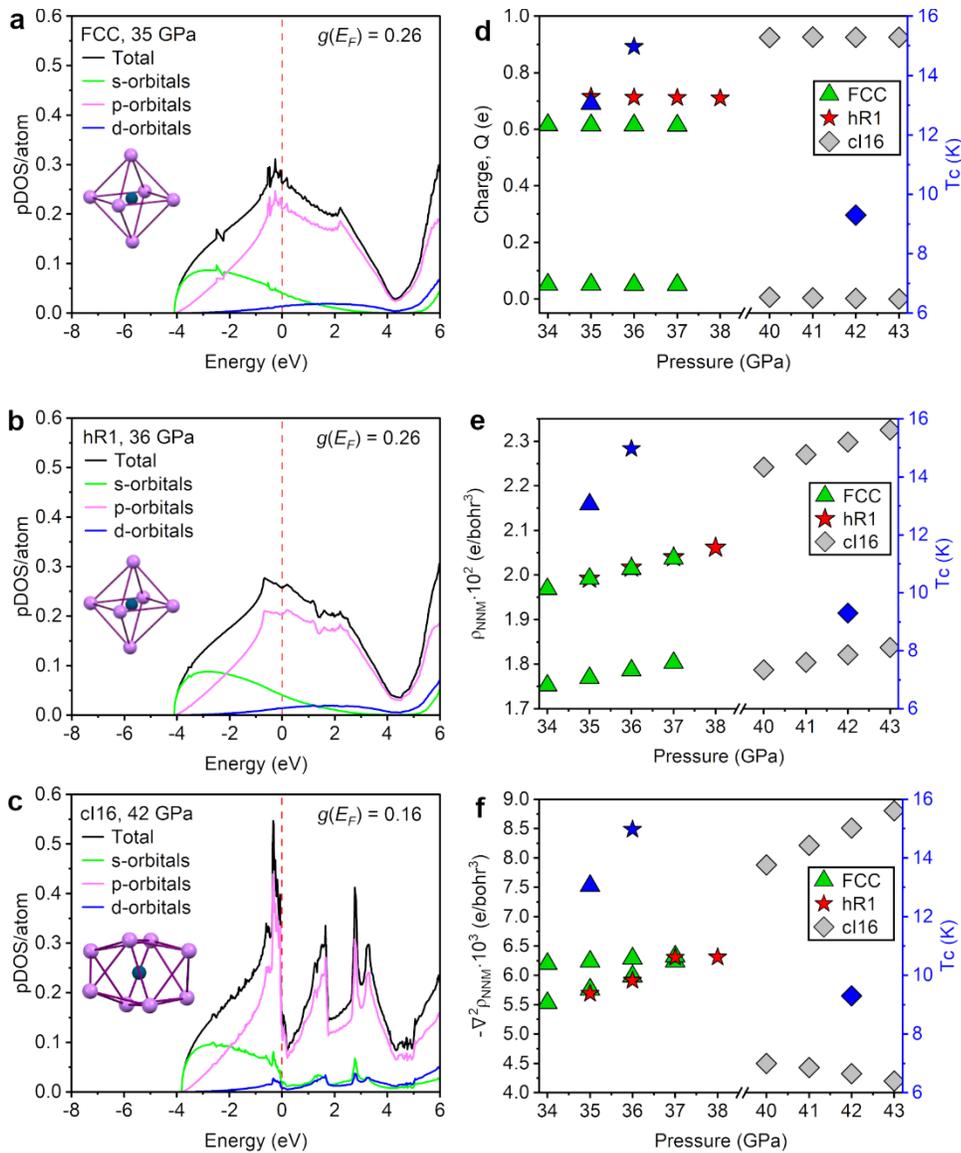

**Figure 4.** Projected DOS (pDOS) and values of the density of states at the Fermi level ($g(E_F)$) of Li FCC, hR1 and cI16 at the indicated pressures (**a-c**). The insets in (a-c) illustrate the polyhedra (in pink) that surround the ELF loci for the primary NNM (blue spheres) in these phases. The Fermi level is set at the zero of the energy axes. Values of charges Q (**d**), $\rho_{NNM}$ (**e**) and $-\nabla^2\rho_{NNM}$ (**f**) calculated at the NNM for the three Li phases as a function of pressure and compared to the calculated $T$c values (reported in blue).

Both QTAIM and ELF calculate two non-equivalent basins in FCC and cI16. However, in both cases, the secondary basin is very small, accounting for a total integrated charge, Q, close to zero, and therefore it is not considered in the discussion below. The absolute value of Q in the primary basin increases in the series FCC → hR1 → cI16 and slightly upon compression (Figure 4d); but with a marked difference from hR1 to cI16. Similarly, the value of the electron density at the NNM, $\rho_{NNM}$, (Figure 4e) increases almost linearly with pressure within the two families of structures. The value of $\rho_{NNM}$ measured in the principal basin of FCC coincides with that of hR1 at the same pressure points. Then, $\rho_{NNM}$ increases abruptly for the transition to cI16. Interestingly, the degree of concentration of the electron density at the NNM, quantified by the curvature, $-\nabla^2\rho_{NNM}$, (Figure 4f), become equivalent for FCC and hR1 near the pressure of phase transition, suggesting that the two optimized geometries are nearly indistinguishable. Meanwhile, in cI16, only the curvature of the main NNM becomes more pronounced, and this curvature decreases for the second basin. Overall, these topological descriptors find the charge density within cI16, which corresponds to the intermetallic interactions, as very concentrated and localized.

Why is cI16 so different? The origin of this difference might be related to the atomic packing and structural change in response to compression (Figure 4a-c and S6a-c), which can be related to the distortion of a BCC lattice [5]. The local geometry of the Li atoms surrounding the main NNM in FCC has the shape of a regular octahedron, with the NNM at its center (Figure 4a, inset). In hR1, the local geometry is almost identical (Figure 4b, inset), but the $Li_6$ octahedron surrounding the NNM is now distorted, with the length of six Li-Li edges equal to 2.41 Å and the other six equal to 2.30 Å (at 40 GPa). With the increase of pressure, more lithium atoms will inevitably move closer to each other and to the NNM (eight atoms in total, see inset in Figure 4c), forcing the electron density to reorganize so as to minimize repulsions and maximize bonding stabilization, similar to what is observed in high pressure phases of sodium [30]. Consequently, in cI16, the main NNM is at the center of a $Li_8$ cluster having the shape of a dodecadeltahedron (Figure 4c, inset). Therefore, a larger number of hybrid sp orbitals on the lithium atoms overlap in cI16, as compared to hR1 or FCC, to form the NNM in the electron density. Highly concentrated and localized electrons (Figure 4d-f) are less polarizable (less free to move), causing the opening of a pseudo-gap (Figure 4c), and hampering the electron-phonon coupling (Figure 2c).

**Conclusions**

In summary, we collected experimental data on the phase transitions of lithium using hydrostatic pressure conditions and show that the transition between FCC→ hR1→ cI16 occurs also at low temperatures, when superconductivity is observed in lithium. We further show that small but observable isotope effects are present in the FCC→ hR1 transition and for $^6Li$, the transition to hR1 happens at lower compression both as phase mixture with FCC and as a pure phase. We identified the boundaries of these phases, and our density functional theory calculations reproduced the experimental trend of *T*c in the range of pressure between 35 and 42 GPa, coincident with the occurrence of the phase transitions FCC → hR1 → cI16. We attribute the increase of the *T*c at ~35 – 36 GPa to the presence of soft phonon modes in the FCC and hR1 phases. The subsequent drop of *T*c upon transition to cI16 in the range of ~ 40 GPa is explained by a decrease in the electron-phonon coupling, which is sided by a large localization and concentration of the electron density in the Li-Li interactions, as response to a drastic structural change.

**Acknowledgements**
The authors are grateful for the experimental assistance provided by Tushar Bhowmick, Mason Burden, Adam Dockery, Audrey Glende, Alice Leppert, Tessa McNamee, Fatemeh Safari, Saveez Saffarian, and Julia St. Andre during sample preparation and XRD data collection. Funding for this research is provided by the National Science Foundation under Awards PHY-2020249 (S. R.) and DMR-2136038 (F. B.). Calculations were performed at the Center for Computational Research at SUNY Buffalo (http://hdl.handle.net/10477/79221). The experimental research at the University of Utah received support from the National Science Foundation Division of Materials Research Award No. 2132692. This work was also supported by the U.S. Department of Energy (DOE) Office of Science, Fusion Energy Sciences funding award entitled *High Energy Density Quantum Matter,* Award No DE-SC0020340. Travel funds for student participation in experiments at Argonne National Laboratory were provided in part by the Undergraduate Research Opportunities Program of the University of Utah, the National Science Foundation (NSF) award #1950409, and the University of Utah Physics & Astronomy Summer Undergraduate Research Program (SURP). The experimental work took place at HPCAT (Sector 16), Advanced Photon Source (APS), Argonne National Laboratory. HPCAT operations are supported by DOE-NNSA's Office of Experimental Sciences. The APS is a DOE Office of Science User Facility operated for the DOE Office of Science by Argonne National Laboratory under contract DE-AC02-06CH11357.

**Supplemental Material**
The Supplemental Material provides detailed description of the experimental and theoretical study.

# Supplemental Material
# Phase Boundaries, Isotope Effect and Superconductivity of Lithium Under Hydrostatic Conditions

## S1- Experimental Details

Symmetric diamond anvil cells having culet sizes of 350 μm were used for generation of high pressure (Figure S1a). Stainless steel or rhenium was used as a gasket material. To prevent reaction of lithium with the diamonds, we coated the diamonds with a thin layer of $Al_2O_3$ (~15$nm$) using atomic layer deposition. Isotopically enriched samples of [7]Li (with 98% enrichment level from Sigma-Aldrich) and [6]Li rich (95.6%) Sigma-Aldrich were loaded in the pressure chamber of the DAC, together with ruby chips and an Au piece for pressure calibration, in an argon glove box with oxygen and water levels kept below 0.1 ppm. Helium was loaded as a pressure transmitting medium in the gas loading facility of Sector 13-, GSECARS, of the Advanced Photon Source, Argonne National Laboratory. Pressure was determined using fluorescence of ruby and was confirmed by the equation of state of Au (Figure S1a). High-pressure diffraction data were collected at 16-ID-B beamline, HPCAT of the Advanced Photon Source (APS), Argonne National Laboratory using x-ray wavelengths of 0.4066 Å or 0.4246 Å. The DACs were rotated by 20° at a rate of 0.25°/s and the data were integrated in 83 s exposure time. A double membrane assembly was used to change the pressure. After initial pressurization within the BCC structure at room temperature, the DACs were cooled down to low temperatures and kept below 100 K, similar to the conditions where superconductivity data were collected previously, to prevent reaction of lithium with diamonds. Diffraction data was collected between 10 – 80 K and at pressure intervals of ~1 GPa and the XRD measurements were analyzed using DIOPTAS software [1] (Figure S1b-c). Data was collected from several spots within the pressure cell and the background from regions without sample was collected to exclude the peaks not originating from the sample.

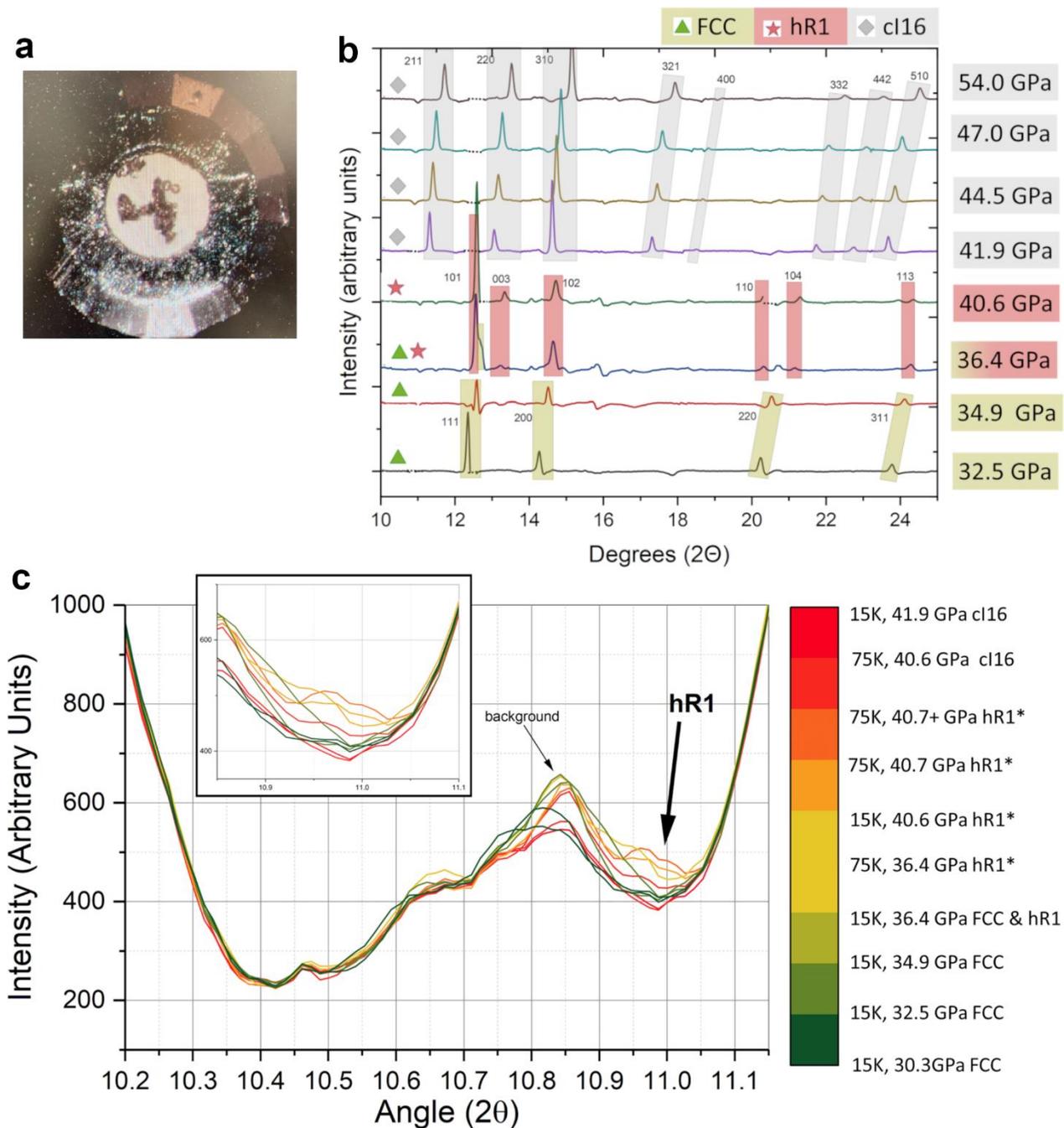

**Figure S1.** (**a**) A Micrograph of lithium sample pressurized in helium pressure medium in a diamond anvil cell. (**b**) The PXRD pattern of $^7$Li under compression at 15 K. (**c**) Details of the PXRD pattern of $^7$Li for different phases (FCC, hR1, cI16) at various pressures and temperatures ($T$ = 15 – 75 K, $P$ = 30.3 – 41.9 GPa), indicating a new peak at around ~10.95 ° in 2θ, which appears exclusively in the hR1 phase where the sample is present. The peak at 10.85 ° is a background peak that is present also without the sample. hR1*: Since equation of states of Li in hR1 phase is not established, the pressures in pure hR1 phase are estimated by interpolation.

## S2 - Comparison with the superconductivity data

To ensure a meaningful comparison of superconductivity data and the identification of phase transition boundaries between the two isotopes, it is crucial to create conditions as similar as possible across different measurements. In this study, we compare both our experimental and theoretical findings with those reported in reference [2], where helium serves as the pressure-transmitting medium (PTM). The variations in hydrostaticity and the methods used to determine the superconducting critical temperature, $T_c$, can lead to significant differences in the observed values. For instance, Figure S2 presents the results of measurements conducted in Ref. [2] using helium as a PTM, where superconductivity is determined by the AC magnetic susceptibility method. These results are compared with those in the study by Struzhkin *et al.* [3], where no PTM is used, and measurements are performed partly using AC magnetic susceptibility at lower pressures and using the four-probe electrical resistivity technique at higher pressures. Major discrepancies between the two studies are primarily observed when different methods are employed. This disparity can be understood considering that electrical resistivity measurements are path-dependent, and even if a small part of the sample exhibits superconductivity, these measurements would indicate a drop in resistance. A large pressure gradient can be present within the pressure cell, which can be significant at higher pressures and under non-hydrostatic conditions used in these types of measurements. Therefore, electrical resistivity measurements, which rely on the onset of a drop in electrical resistance, are biased toward the region of the sample with the highest $T_c$ and tend to overestimate $T_c$. On the other hand, AC magnetic susceptibility detects the exclusion of the magnetic field from the sample and depends on the surface of the superconducting portion. Therefore, these measurements are more prone to underestimating $T_c$. Figure S2 illustrates such differences.

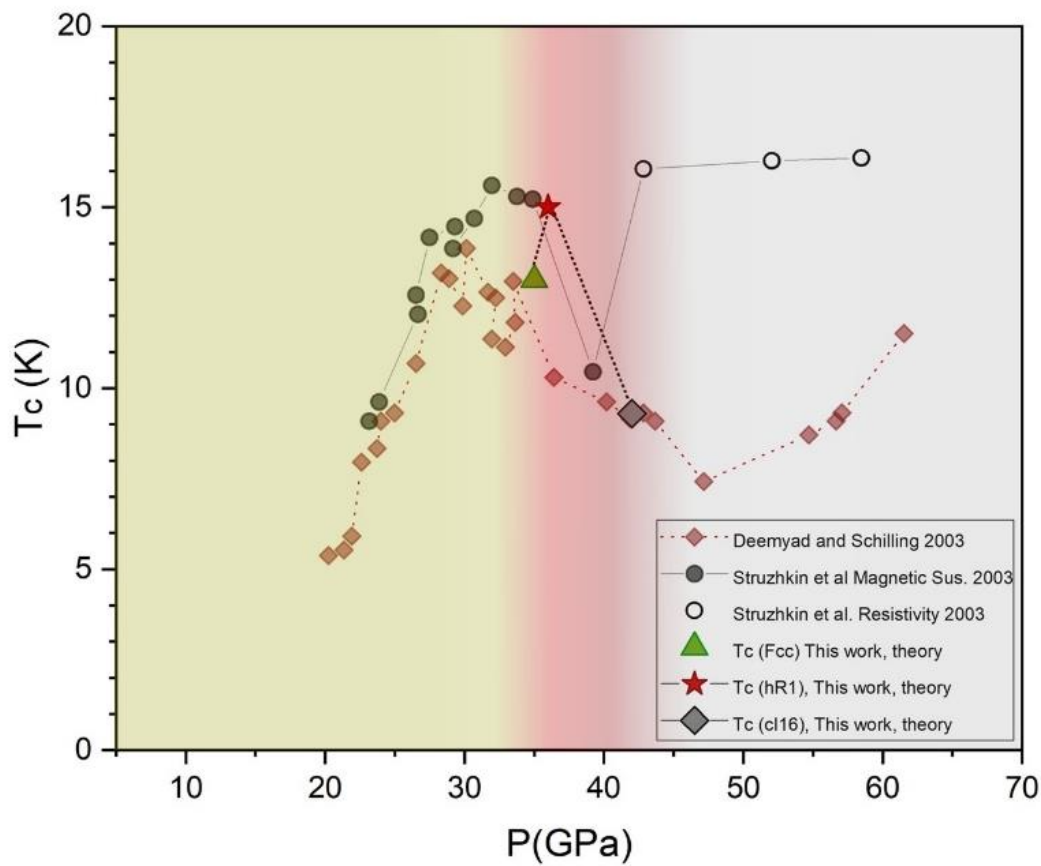

**Figure S2**. Superconducting phase diagram of [7]Li reported in two different studies, along with the theoretical calculations presented here. Red diamonds and closed black circles represent measurements of superconductivity using AC magnetic susceptibility in hydrostatic (helium as PTM) and non-hydrostatic (no PTM) conditions, as conducted in two separate studies by Deemyad and Schilling, and Struzhkin *et al.*, respectively (Ref. [2,3]). The open circles depict resistivity measurements by Struzhkin *et al.* under non-hydrostatic conditions in the same study. There are notable variations between the two studies, particularly when different techniques are employed.

### S3 - Computational Details and Theoretical Results

We performed periodic Density Functional Theory (DFT) calculations using the plane-wave based Vienna *Ab-Initio* Simulation Package (VASP, version 6.2.1) [4] and Quantum Espresso (QE, version 7.2) [5] codes, along with the atomic orbital-based Crystal17 code [6].

For VASP, the PBE [7,8] and the R$^2$SCAN-L [9] exchange-correlation functional were employed for the geometry optimizations and calculations of the electronic properties. The projected augmented wave (PAW) method [10] was used, in combination with the POTCAR *PAW_PBE Li*, which treats explicitly the Li $2s^1$ states, and a cutoff of 550 eV. At these pressures, the 1s states are not interacting, and well described by a pseudopotential [11,12], so it is not necessary to treat them explicitly in the calculations. The *k*-point mesh was generated using the Γ-centered Monkhorst-Pack scheme [13], and the number of divisions along each reciprocal lattice vector was selected so that the product of this number with the real lattice constant was greater than or equal to 80 Å. The topological analysis of the electron density, based on the Quantum Theory of Atoms in Molecules (QTAIM), [14] was performed using the Critic2 [15] code. Phonons in the harmonic approximation were determined with the Phonopy package [16] using supercells equal to 4x4x4 for FCC and to 5x5x2 for hR1, based on the conventional unit cells.

For QE, the PBE [7,8] exchange-correlation functional was employed in combination with the ultrasoft pseudopotential Li.pbe-n-van.UPF. The FCC, hR1 and cI16 geometries were optimized at 35 GPa, 36 GPa and 42 GPa, respectively. A 100 Ry energy cutoff was employed along with a charge-density cutoff of 800 Ry for the valence electrons. To ensure the convergency of the energy and of the electron-phonon coupling constant within a Gaussian broadening of 0.005, a *k*-mesh grid of 40x40x40, 50x50x30 and 20x20x20 and *q*-point grids of 8x8x8, 10x10x6, and 4x4x4 were used on the conventional unit cells of FCC, hR1 and cI16, respectively. The superconducting critical temperature (*T*c) was estimated using the Allen-Dynes modified McMillan equation [17].

$$k_B T_c = \frac{\omega_{ln}}{1.2} exp\left[\frac{-1.04(1+\lambda)}{\lambda - \mu^*(1+0.62\lambda)}\right], \quad (1)$$

in which the effective Coulomb potential, $\mu^*$, was set equal to 0.17, as calculated and used by Borinaga *et al.* [18]. The terms $\lambda$ and $\omega_{ln}$ were calculated from the isotropic version of the Eliashberg function and represent the electron-phonon coupling parameter and the logarithmic average phonon frequency, calculated as:

$$\lambda = 2\int_0^\infty \frac{\alpha^2 F(\omega)}{\omega} d\omega, \quad (3)$$

$$\omega_{ln} = exp\left[\frac{2}{\lambda}\int_0^\infty \frac{d\omega}{\omega}\alpha^2 F(\omega) ln\omega\right]. \quad (4)$$

Here, $\alpha^2 F(\omega)$ is the isotropic version of the Eliashberg function, which is dependent on the density of states (DOS), the phonon density of states (PHDOS), and the electron phonon matrix elements. [19]

For Crystal17, and the calculation of the structure factors, $F_{hkl}$, we employed the HSE06 [20–22] exchange-correlation functional on the hR1 geometry optimized with VASP (PBE). The k-point grid was generated with the Monkhorst-Pack method, using a shrinking factor of 32 along the reciprocal lattice vectors (32x32x32 grid in k-space), a smearing of 0.001 Ha, and a convergence threshold of the total energy equal to $10^{-8}$ Ha. For this calculation, we extended and adapted the TZVP-rev2 (https://www.crystal.unito.it/Basis_Sets/lithium.html#Li_pob_TZVP_2012) basis-set of lithium as follows:

```
3 7
0 0 6 2.0 1.0
  6269.2628010      0.00020540968826
  940.31612431      0.00159165540890
  214.22107528      0.00828698297070
```

```
 60.759840184       0.03385637424900
 19.915152032       0.11103225876000
  7.3171509797      0.27449383329000
0 0 2 1.0 1.0
  2.9724674216      0.23792456411000
  1.2639852314      0.30765411924000
0 0 1 0.0 1.0
  0.5025516200      1.00000000000000
0 0 1 0.0 1.0
  0.075     1.00000000000000
0 2 1 0.0 1.0
  1.19      1.00000000000000
0 2 1 0.0 1.0
  0.190     1.00000000000000
0 3 1 0. 1.
  0.30 1.
```

**Table S1.** Theoretical $F_{hkl}$ for the (100) reflection, calculated with Crystal17 for the Li hR1 geometry, optimized with VASP at 40 GPa, and for distorted geometries generated by varying the γ cell parameter (γ = 120° in the optimized geometry).

| γ | 120° | 121° | 118° |
|---|---|---|---|
| 2θ | 11.38 | 11.39 | 10.93 |
| $F_{hkl}$ | 0.0 | 1.12E-05 | 1.05E-05 |

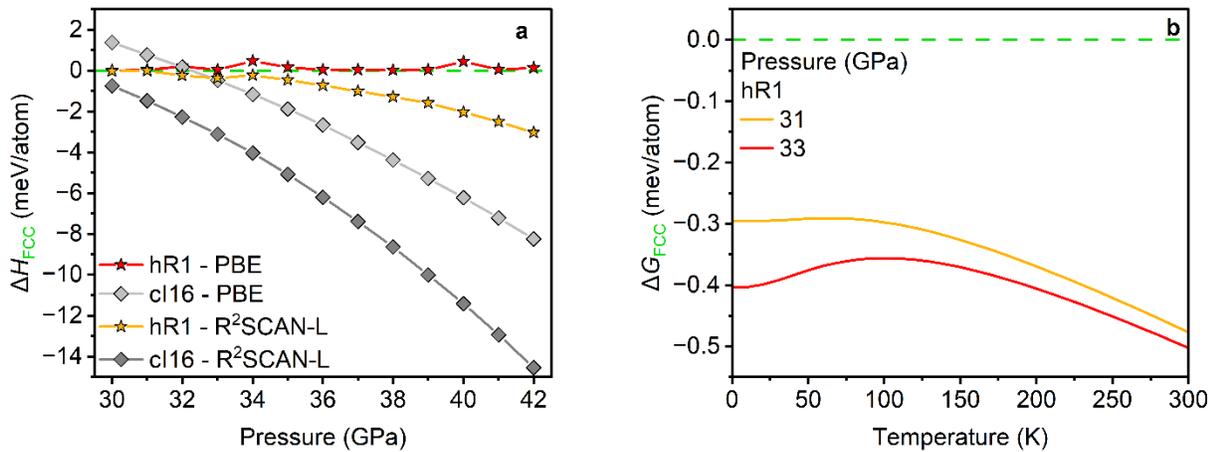

**Figure S3.** (**a**) Pressure dependent enthalpy differences for hR1 and cI16 relative to the FCC phase, calculated with the PBE and R²SCAN-L functionals with VASP. (**b**) Gibbs free energy difference calculated at 31 and 33 GPa for hR1 relative to the FCC phase calculated with the PBE functional.

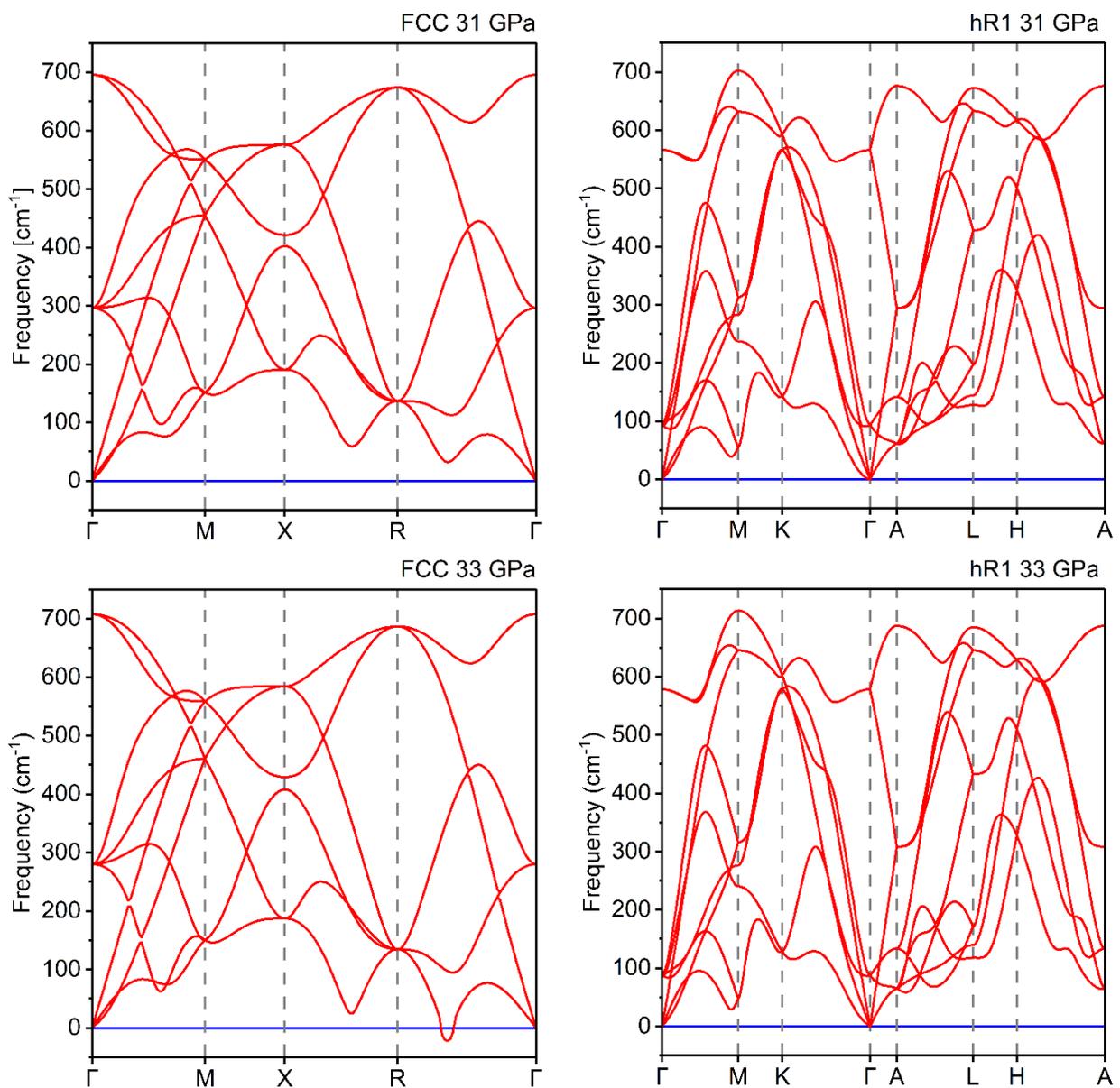

**Figure S4.** Phonon band structures of FCC and hR1 calculated with PBE at pressures of 31 and 33 GPa.

## S4 - Quantum Ionic Fluctuations

Quantum ionic effects on the FCC and hR1 phases of lithium were studied using the Stochastic Self-Consistent Harmonic Approximation (SSCHA) [23] in conjunction with machine learning interatomic potentials trained using the MLIP package [24].

The anharmonic phonon spectra were first calculated on a 4x4x4 supercell (64 atoms) for the FCC phase, and a 3x3x3 supercell (81 atoms) for the hR1 phase, both using up to 1000 configurations in a range of pressures close to 38 GPa. The DFT calculations for the SSCHA have been performed through the plane wave based code Quantum Espresso [5]. The integrations over the Brillouin zone for the FCC and hR1 phases were performed with a first-order Methfessel-Paxton smearing with broadenings of 0.108 and 0.082 eV, 8x8x8 and 5x5x3 k-point grids, respectively, and a kinetic energy cutoff of 1500 eV. PAW pseudopotentials were employed with the PBE exchange-correlation functional [7,8], and the three electrons of lithium were treated explicitly, along with the description up to the 2p orbitals for the lithium.

The configurations used for the calculation of the SSCHA anharmonic phonons in the aforementioned FCC and hR1 supercells were then employed to train a Level 10 Momentum Tensor Potential (MTP) with a cutoff radius of 9 Angstrom using the MLIP package. The anharmonic phonon spectra shown in this work were calculated through the SSCHA at different temperatures within the bubble approximation (Eq. 59, Ref. [23]). This was done on 6x6x6 and 6x6x4 supercells of the FCC and hR1 structures, respectively, using up to 1000 configurations. Energies, forces, and stresses for each configuration were obtained through the MTP potential. Additional configurations were fed to the potential using an active learning procedure based on the Maxvol Algorithm [24] with an extrapolation grade of 2.

The resulting FCC and hR1 phonon spectra calculated at 38 GPa and at 0 K are shown in Figure S5, along with the harmonic phonons at similar pressures. The quantum anharmonic behavior of lithium introduces substantial renormalization for the lowest optical branches of both structures. The FCC phase is found to be unstable at this pressure and temperature. The anharmonic dynamic instability is driven by a soft mode along the Γ – M path and by a mode along the R – Γ path, which was partially captured also by the harmonic approximation at lower pressure (Figure S4).

The hR1 structure is instead dynamically stable, featuring a soft mode on the Γ – A path (Figure S5), analogous to the prediction with the harmonic approximation (Figure S4 and Figure 3). However, in contrast to the simpler harmonic calculations, the anharmonic effects sustain the phonons around the M high symmetry point, and introduce a red shift along the Γ – A – L path, highlighting the presence of substantial quantum anharmonicity (Figure S5). At 0 K the hR1 structure remains stable only in a short range of pressures between 38 and 39 GPa. Increasing the temperature, however, the structure was found to be unstable due to the same soft mode along the Γ – A path. An apparent imaginary mode along Γ – K can be noticed in hR1 (Figure S5b). This mode is not a dynamical instability, but an interpolation error, which could not be corrected using a finer phonon grid due to the high computational load required to compute the anharmonic phonon spectra.

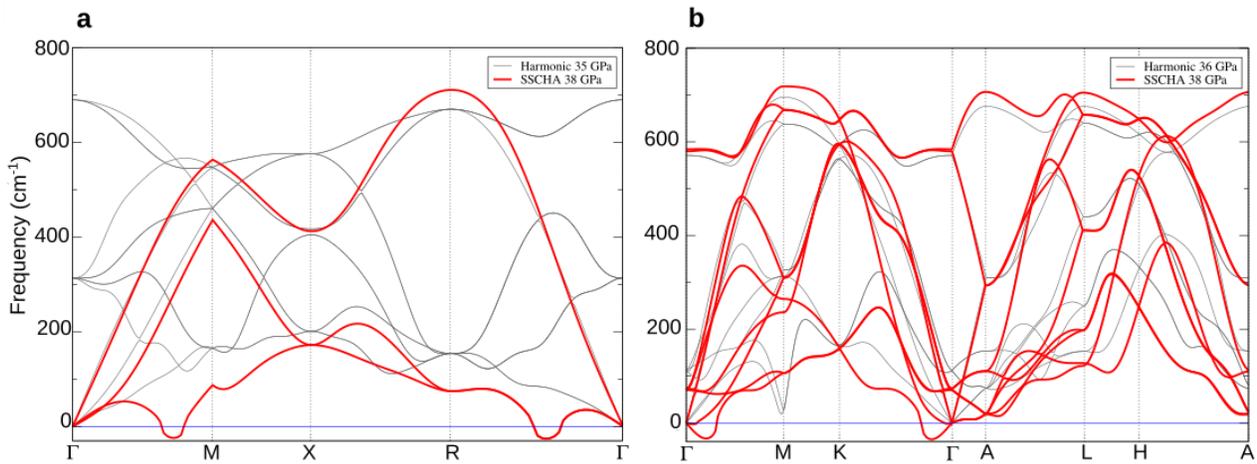

**Figure S5.** Anharmonic phonon spectra (red) for the primitive FCC unit cell overlapped with the harmonic phonon spectra (gray) of the conventional FCC unit cell (**a**) and anharmonic (red) and harmonic (gray) phonon spectra for the conventional hR1 (**b**) unit cell. Both anharmonic phonon spectra are calculated through the SSCHA at 38 GPa. The imaginary frequency near Γ for the hR1 structure is related to an interpolation error, while FCC is dynamically unstable.

**Table S2.** Critical Temperature ($T_c$), electron-phonon coupling constant ($\lambda$) and the logarithmic average phonon frequency factor ($\omega_{ln}$) calculated with QE for the following lithium phases: FCC at 35 GPa, hR1 at 36 GPa and cI16 at 42 GPa (harmonic approximation).

| Phase | Space Group | Pressure (GPa) | $T_c{}^a$ (K) | $\lambda$ | $\omega_{ln}$ (cm$^{-1}$) |
|---|---|---|---|---|---|
| FCC | Fm-3m | 35 | 13.0 | 1.03 | 247.8 |
| hR1 | R-3m | 36 | 15.0 | 1.15 | 226.9 |
| cI16 | I-43d | 42 | 9.3 | 0.93 | 221.1 |

[a] $\mu^* = 0.17$ [13].

## S5 - Overview of the QTAIM and the ELF

The topological analysis of the electron density is based on the Quantum Theory of Atoms in Molecules (QTAIM), by Richard Bader. [14] The purpose of this analysis is to retrieve chemically relevant information from the electron density, ρ(**r**). At the core of QTAIM is the partition of ρ(**r**) into discrete atomic entities. The atomic boundaries are defined by the surface zero-flux condition of the electron density calculated for each atom in the system [25]. The region of space defined by the zero-flux condition is called an atomic basin. Each basin is a well-defined physically meaningful entity that satisfies the virial theorem [26]. Atomic basins have well-defined shapes (usually non-spherical) and volumes, and they can be used to integrate electronic charges, Q [e], and other properties [26,27]. The topological analysis of ρ(**r**) reveals also critical points that can correspond to atomic positions (all curvatures are negative, *i.e.*, a maximum), chemical bonds (two negative curvatures and one positive curvature, *i.e.*, saddle-points), or other special points of ρ(**r**) [28].

Critical points related with chemical interactions, as well as metallic bonds, can be described and characterized by the amount of electron density at their positions, $\rho_{cp}$ [e/bohr$^3$], and by the Laplacian, $\nabla^2\rho_{cp}$ [e/bohr$^5$] [29,30]. Generally, the larger the amount of electron density at the critical point, the larger the electron sharing between atoms. Similarly, the Laplacian quantifies the degree of concentration of $\rho(\mathbf{r})$ at the critical point, where values $\nabla^2\rho_{cp} < 0$, or $-\nabla^2\rho_{cp} > 0$, stands for accumulation of electron density.

A non-nuclear attractor, or non-nuclear maximum (NNM), is a critical point corresponding to a maximum of the $\rho(\mathbf{r})$ that does not coincide with an atomic position. Richard Bader and Carlo Gatti were the first to notice that alkali metals are characterized by NNM of the electron densities. It was then established that this accumulation is often generated by the chemical interactions between electropositive metals, such as alkali and alkaline earth, but also scandium and aluminum [30].

The electron localization function (ELF) is defined by the probability of finding two electrons having the same spin in a region of space, and therefore, it is a useful method to visualize Fermi holes and highlight locations characterized by large Pauli repulsion. The values of ELF span from 0.5, for regions of space possessing the same features of a free electron gas, to 1, for locations characterized by Fermi holes having very low curvature (large exclusion of electrons having the same spin). Therefore, regions of space having values of ELF close to 1, can be associated with either isolated electrons or electron pairs having opposite spin [31].

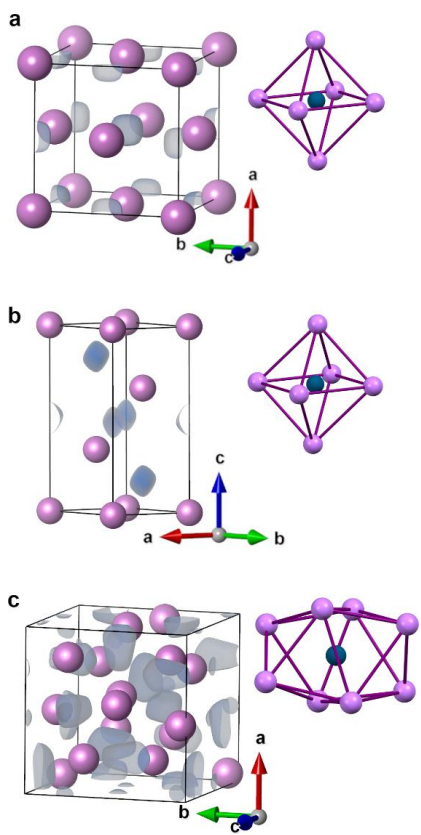

**Figure S6**. ELF's isosurfaces (values = 0.8) and lithium clusters (in pink) surrounding the primary NNM (in blue) in FCC (**a**), hR1 (**b**) and cI16 (**c**).